\documentclass[aps,prc,twocolumn,showpacs]{revtex4}
\usepackage{graphicx,amsfonts,amssymb,amsbsy}
\usepackage{pstricks}
\usepackage{amsmath }

\begin{document}

\title{Properties of strongly magnetized ultradense matter and their  imprints on magnetar pulsations}
\author{C. V\'asquez Flores$^1$,  L. B. Castro$^2$,  G. Lugones$^1$}

\affiliation{$^1$ Centro de Ci\^{e}ncias Naturais e Humanas, Universidade Federal do ABC, \\ Av. dos Estados 5001, CEP 09210-580, Santo Andr\'{e}, SP, Brazil}
\affiliation{$^2$ Departamento de F\'{\i}sica, Universidade Federal do Maranh\~{a}o, \\Campus Universit\'{a}rio do Bacanga, CEP 65080-805, S\~{a}o Lu\'{i}s, Maranh\~{a}o, Brazil}

\begin{abstract}
We investigate the effect of strong magnetic fields on the adiabatic radial oscillations of hadronic stars. We describe magnetized hadronic matter within the framework of the relativistic nonlinear Walecka model and integrate the equations of relativistic radial oscillations to determine the fundamental pulsation mode.  We consider that the magnetic field increases, in a density dependent way, from the surface, where it has a typical magnetar value of $10^{15}$ G, to the interior of the star where it can be as large as $3 \times 10^{18}$ G. 
We show that magnetic fields of the order of $10^{18}$ G at the stellar core produce a significant change in the frequency of neutron star pulsations with respect to unmagnetized objects. If radial pulsations are excited in magnetar flares, they can leave an imprint in the flare lightcurves and open a new window for the study of highly magnetized ultradense matter.
\end{abstract}

\email{cesarovfsky@gmail.com}

\email{lrb.castro@ufma.br}

\email{german.lugones@ufabc.edu.br}


\maketitle

\section{Introduction}
\label{sec:intro}

Compact stars have a large number of pulsation modes that have been extensively studied since the seminal work of Chandrasekhar on radial oscillations \cite{APJ140:417:1964,PRL12:114:1964}. In general, these modes are very difficult to observe in the electromagnetic spectrum; therefore most efforts have concentrated on gravitational wave asteroseismology in order to characterise the frequency and damping times of the modes that emit gravitational radiation. In particular, various works focused on the oscillatory properties of  pure hadronic stars, hybrid stars and strange quark stars trying to find signatures of the equation of state of high density neutron star matter  (see \cite{AA325:217:1997,IJMPD07:29:1998,AA366:565:2001,APJ579:374:2002,PRD82:063006:2010,EL105:39001:2014} and references therein).

More recently, compact star oscillations have attracted the attention in the context of Soft Gamma ray Repeaters (SGRs), which are persistent X-ray emitters that sporadically emit short bursts of soft $\gamma$-rays. In the quiescent state, SGRs have an X-ray luminosity of $\sim 10^{35}$ erg/s, while during the short $\gamma$-bursts they release up to $10^{42}$ erg/s in episodes of about 0.1 s.   Exceptionally, some  of them have emitted very energetic giant flares which commenced with brief $\gamma$-ray spikes of $\sim 0.2$ s, followed by tails lasting hundreds of seconds. Hard spectra (up to 1 MeV) were observed during the spike and the hard X-ray emission of the tail gradually faded modulated at the neutron star (NS) rotation period. The analysis of X-ray data of the tails of the giant flares of SGR 0526-66, SGR 1900+14 and SGR 1806-20 revealed the presence of quasi-periodic oscillations (QPOs)  with frequencies ranging from $\sim$ 18 to 1840  Hz \cite{APJ628:L53:2005,APJ632:L111:2005,AA528:A45:2011}. There are also candidate QPOs at higher frequencies up to $\sim 4$ kHz in other bursts but with lower statistical significance \cite{ElMezeini2010}; in fact, according to a more recent analysis only one burst shows a marginally significant signal at a frequency of around 3706 Hz \cite{Huppenkothen2013}. 
 
Several characteristics of SGRs are usually explained in terms of the \textit{magnetar} model, assuming that the object is a neutron star with an unusually strong magnetic field ($B \sim 10^{15} $ G) \cite{Woods2006}.  In particular, giant flares are associated to catastrophic rearrangements of the magnetic field. Such violent phenomena are expected to excite a variety of oscillation modes in the stellar crust and core. In fact,  recent studies have accounted for magnetic coupling between the crust and the core, and associate QPOs to global magneto-elastic oscillations of highly magnetized neutron stars \cite{Levin2007,CerdaDuran2009,Colaiuda2012,Gabler2014}. There has also been interest in the possible excitation of low order $f$-modes because of their strong coupling to  potentially detectable gravitational radiation \cite{Levin2011}. 

In the present paper we focus on radial oscillations of neutron stars permeated by ultra-strong magnetic fields. These modes might be relevant within the magnetar model because they could be excited during the violent events associated with gamma flares. Since they have higher frequencies than the already known QPOs, they cannot be directly linked to them at present. However, it is relevant to know all the variety of pulsation modes of strongly magnetized neutron stars because the number of observations is still small and new features could emerge in future flares' data. 
On the other hand, in the case of rotating objects we can expect  some amount of gravitational radiation from even the lowest ($l = 0$) quasi-radial mode \cite{Stergioulas2003,Passamonti2006} making them potentially relevant for  gravitational wave astronomy.

\section{Equations of state}
\label{sec2}

\subsection{Hadronic phase under a magnetic field}
\label{sec:A}

In this section we present an overview of the hadronic equations of state (EOS) used in this work. We describe hadronic matter within the framework of the relativistic non-linear Walecka (NLW) model \cite{WALECKA1986}. In this model we employ a field-theoretical approach in which the baryons interact via the exchange of $\sigma-\omega-\rho$ mesons in the presence of a magnetic field $B$ along the $z-$axis. The total lagrangian density reads:
\begin{equation}\label{lt}
    \mathcal{L}_{H}=\sum_{b}\mathcal{L}_{b}+\mathcal{L}_{m}+\sum_{l}\mathcal{L}_{l}+\mathcal{L}_B\,.
\end{equation}
where $\mathcal{L}_{b}$, $\mathcal{L}_{m}$, $\mathcal{L}_{l}$ and $\mathcal{L}_{B}$ are the baryons, mesons, leptons and electromagnetic field Lagrangians, respectively, and are given by
\begin{eqnarray}
    \mathcal{L}_{b} &=& \overline{\psi}_{b}\left(i\gamma_{\mu}\partial^{\mu}-q_{b}\gamma_{\mu}A^{\mu}-m_{b}+g_{\sigma b}\sigma \right. \nonumber \\
    && \left. -g_{\omega b}\gamma_{\mu}\omega^{\mu}-g_{\rho b}\tau_{3b}\gamma_{\mu}\rho^{\mu}\right)\psi_{b}\,,  \\
   \mathcal{L}_{m} &=& \tfrac{1}{2}(\partial_{\mu}\sigma\partial^{\mu}\sigma-m_{\sigma}^{2}\sigma^{2})-U(\sigma)+
    \tfrac{1}{2}m_{\omega}^{2}\omega_{\mu}\omega^{\mu} \nonumber \\
   && -\tfrac{1}{4}\Omega_{\mu\nu}\Omega^{\mu\nu}+
    \tfrac{1}{2}m_{\rho}^{2}\vec{\rho}_{\mu}\cdot\vec{\rho}_{\mu}-\tfrac{1}{4}P^{\mu\nu}P_{\mu\nu} \,,   \\
    \mathcal{L}_{l} &=& \overline{\psi}_{l}\left(i\gamma_{\mu}\partial^{\mu}-q_{l}\gamma_{\mu}A^{\mu}-m_{l}\right)\psi_{l} \,,   \\
    \mathcal{L}_{B} &=& -\tfrac{1}{4}F^{\mu\nu}F_{\mu\nu} \,.
\end{eqnarray}
where he $b$-sum runs over the baryonic octet $b\equiv N~(p,~n),~\Lambda,~\Sigma^{\pm,0},~\Xi^{-,0}$, $\psi_{b}$ is the corresponding baryon Dirac field, whose interactions are mediated by the $\sigma$ scalar, $\omega_{\mu}$ isoscalar-vector and $\rho_{\mu}$ isovector-vector meson fields. The baryon charge, baryon mass and isospin projection are denoted by $q_{b}$, $m_{b}$ and $\tau_{3b}$, respectively, and the masses of the mesons are $ m_{\sigma}= 512~$ MeV, $m_{\omega}=783~$MeV and $m_{\rho}=770~$MeV. The strong interaction couplings of the nucleons with the meson fields are denoted by $g_{\sigma N}=8.910$, $g_{\omega N}=10.610$ and $g_{\rho N}=8.196$. We consider that  the couplings of the hyperons with the meson fields are fractions of those of the nucleons, defining $g_{iH}=X_{iH}g_{iN}$, where the values of $X_{iH}$ are chosen as $X_{\sigma H}=0.700$ and $X_{\omega H}=X_{\rho H}=0.783$ \cite{GLENDENNING2000}.
 The term $U(\sigma)=\frac{1}{3}\,bm_{n}(g_{\sigma N}\sigma)^{3}-\frac{1}{4}\,c(g_{\sigma N}\sigma)^{4}$ denotes the scalar self-interactions \cite{NPA292:413:1977,PLB114:392:1982,AJ293:470:1985}, with
 $c=-0.001070$ and $b=0.002947$. The mesonic and electromagnetic field tensors are given by their usual expressions $\Omega_{\mu\nu}=\partial_{\mu}\omega_{\nu}-\partial_{\nu}\omega_{\mu}$, ${\bf P}_{\mu\nu}=\partial_{\mu}\vec{\rho}_{\nu}-\partial_{\nu}\vec{\rho}_{\mu}-g_{\rho b}(\vec{\rho}_{\mu}\times\vec{\rho}_{\nu})$ and $F_{\mu\nu}=\partial_{\mu}A_{\nu}-\partial_{\nu}A_{\mu}$. The $l$-sum runs over the two lightest leptons $l\equiv e,\mu$ and $\psi_{l}$ is the lepton Dirac field.
The symmetric nuclear matter properties at saturation density adopted in this work are given by the GM1 parametrization \cite{PRL67:2414:1991}, with compressibility $K=300$ MeV, binding energy $B/A=-16.3$ MeV, symmetry energy $a_{sym}=32.5$ MeV, slope $L=94$ MeV, saturation density $\rho_{0}= 0.153$ fm$^{-3}$ and nucleon mass $m=938$ MeV.

The following equations present the scalar and vector densities for the charged and uncharged baryons \cite{JPG36:115204:2009,PRC89:015805:2014}, respectively:
\begin{eqnarray}
&\label{densities1}\rho_{b}^{s}=\frac{|q_{b}|B \bar{m}_{b} }{2\pi^{2}}\sum_{\nu}^{\nu_{\mathrm{max}}}\sum_{s}\frac{\bar{m}_{b}^{c}}{\sqrt{ \bar{m}_{b}^2    + 2\nu |q_{b}|B}}   \ln\bigg|\frac{k_{F,\nu,s}^{\,b}+E_{F}^{\,b}}{\bar{m}_{b}^{c}} \bigg|,\\
&\label{densities2}\rho_{b}^{v}=\frac{|q_{b}|B}{2\pi^{2}}\sum_{\nu}^{\nu_{\mathrm{max}}}\sum_{s}k_{F,\nu,s}^{\,b},\\
&\label{densities3}\rho_{b}^{s}=\frac{\bar{m}_{b}}{4\pi^{2}}\sum_{s}\bigg[E_{F}^{\,b}k_{F,s}^{\,b}-\bar{m}_{b}^{2}\ln\bigg|\frac{k_{F,s}^{\,b}+E_{F}^{\,b}}{\bar{m}_{b}}\bigg|\bigg] ,\\
&\label{densities4}\rho_{b}^{v}=\frac{1}{2\pi^{2}}\sum_{s}\bigg[\frac{1}{3}(k_{F,s}^{\,b})^{3}\bigg] ,
\end{eqnarray}
 where $\bar{m}_{b}=m_{b}-g_{\sigma}\sigma$ and $\bar{m}_{b}^{c}=\sqrt{ \bar{m}_{b}^2 + 2\nu |q_{b}|B}$.  $\nu=n+\frac{1}{2}-$sgn$(q_{b})\frac{s}{2}=0,1,2,...$ are the Landau levels for the fermions with electric charge $q_{b}$, $s$ is the spin and assumes values
$+1$ for spin up and $-1$ for spin down cases.

The energy spectra for the baryons are given by \cite{JPG35:125201:2008,APJ537:351:2000}:
\begin{eqnarray}\label{energy}
E_{\nu,s}^{\,b}=\sqrt{(k_{z}^{\,b})^{2}+\bar{m}_{b}^{2}+2\nu |q_{b}|B}+g_{\omega b}\omega^{0}+\tau_{3b}g_{\rho b}\rho^{0}\\
E_{s}^{\,b}=\sqrt{(k_{z}^{\,b})^{2}+\bar{m}_{b}^{2}+(k_{\perp}^{\,b})^{2}}+g_{\omega b}\omega^{0}+\tau_{3b}g_{\rho b}\rho^{0},
\end{eqnarray}
where $k_{\perp}^{\,b}=k_{x}^{\,b}+k_{y}^{\,b}$. The Fermi momenta $k_{F,\nu,s}^{\,b}$ of the charged baryons and $k_{F,s}^{\,b}$ of the uncharged baryons and their relationship with the Fermi energies of the charged baryons $E_{F,\nu,s}^{\,b}$ and uncharged baryons $E_{F,s}^{\,b}$ can be written as:
\begin{eqnarray}\label{Momentum}
&&(k_{F,\nu,s}^{\,b})^{2}=(E_{F,\nu,s}^{\,b})^{2}-(\bar{m}_{b}^{c})^{2} \\
&&(k_{F,s}^{\,b})^{2}=(E_{F,s}^{\,b})^{2}-\bar{m}_{b}^{2}.
\end{eqnarray}

For the leptons, the vector density is given by:
\begin{eqnarray}\label{vector_{density_{leptons}}}
\rho_{l}^{v}=\frac{|q_{l}|B}{2\pi^{2}}\sum_{\nu}^{\nu_{\mathrm{max}}}\sum_{s}k_{F,\nu,s}^{\,l}, 
\end{eqnarray}
where $k_{F,\nu,s}^{\,l}$ is the lepton Fermi momentum, which is related to the Fermi energy $E_{F,\nu,s}^{\,l}$ by:
\begin{eqnarray}\label{Momentuml}
(k_{F,\nu,s}^{\,l})^{2}=(E_{F,\nu,s}^{\,l})^{2}-\bar{m}_{l}^{2}\,, \qquad l=e,\mu,
\end{eqnarray}

\noindent with $\bar{m}_{l}=m_{l}^{2}+2\nu |q_{l}|B$.  The summation over the Landau level runs until $\nu_{\mathrm{max}}$; this is the largest value of $\nu$ for which the square of Fermi momenta of the particle is still positive and corresponds to the closest integer, from below to:
\begin{eqnarray}
&&\nu_{\mathrm{max}}=\bigg[\frac{(E_{F}^{\,l})^{2}-m_{l}^{2}}{2|q_{l}|B}\bigg], \quad\mathrm{leptons}\label{ll1}\\
&&\nu_{\mathrm{max}}=\bigg[\frac{(E_{F}^{\,b})^{2}-\bar{m}_{b}^{2}}{2|q_{b}|B}\bigg], \quad\mathrm{charged~baryons}.\label{ll2}
\end{eqnarray}

The chemical potentials of baryons and leptons are:
\begin{eqnarray}\label{chemicalp}
&&\mu_{b}=E_{F}^{\,b}+g_{\omega b}\omega^{0}+\tau_{3b}g_{\rho b}\rho^{0},\\
&&\mu_{l}=E_{F}^{\,l}=\sqrt{(k_{F,\nu,s}^{\,l})^{2}+m_{l}^{2}+2\nu |q_{l}|B}\,.
\end{eqnarray}

From the Lagrangian density~(\ref{lt}), and mean-field approximation, the energy density is given by
\begin{eqnarray}\label{energym}
\varepsilon_{m}= & & \sum_{b}(\varepsilon_{b}^c + \varepsilon_{b}^n) +\tfrac{1}{2}m_{\sigma}\sigma_{0}^{2}\nonumber\\
&& +U(\sigma)+\tfrac{1}{2}m_{\omega}\omega_{0}^{2}+\tfrac{1}{2}m_{\rho}\rho_{0}^{2}\,,
\end{eqnarray}
where the expressions for the energy densities of charged baryons $\varepsilon_{b}^{c}$ and neutral baryons $\varepsilon_{b}^{n}$ are, respectively, given by:
\begin{eqnarray}\label{energy-densities-baryons}
\varepsilon_{b}^{c}& = &\frac{|q_{b}|B}{4\pi^{2}}\sum_{\nu}^{\nu_{\mathrm{max}}}\sum_{s}\bigg[k_{F,\nu,s}^{\,b}E_{F}^{\,b} \nonumber \\
& & + (\bar{m}_{b}^{c})^{2}\ln\bigg|\frac{k_{F,\nu,s}^{\,b}+E_{F}^{\,b}}{\bar{m}_{b}^{c}}\bigg|\bigg],\label{ea1}   \\
\varepsilon_{b}^{n}& = &\frac{1}{4\pi^{2}}\sum_{s}\bigg[\tfrac{1}{2} k_{F,\nu,s}^{\,b}(E_{F}^{\,b})^{3} - \tfrac{1}{4}\bar{m}_{b} \bigg(\bar{m}_{b}k_{F,\nu,s}^{\,b}E_{F}^{\,b} \nonumber \\ 
&& + \bar{m}_{b}^{3}\ln\bigg|\frac{E_{F}^{\,b}+k_{F,\nu,s}^{\,b}}{\bar{m}_{b}}\bigg|\bigg)\bigg]\, . \label{ea2}
\end{eqnarray}

\noindent The expression for the energy density of leptons $\varepsilon_{l}$ reads
\begin{equation}
    \varepsilon_{l}= \frac{|q_{l}|B}{4\pi^{2}}\sum_{l}\sum_{\nu}^{\nu_{\mathrm{max}}}\sum_{s}\bigg[k_{F,\nu,s}^{\,l}E_{F}^{\,l}+
\bar{m}_{l}^{2}\ln\bigg|\frac{k_{F,\nu,s}^{\,l}+E_{F}^{\,l}}{\bar{m}_{l}}\bigg|\bigg]\,. \label{ea3}
\end{equation}

The pressures of baryons and leptons are:
\begin{eqnarray}\label{pressurem}
P_{m}&=&\mu_{n}\sum_{b}\rho_{b}^{v}-\varepsilon_{m}, \\
P_{l}&=&\sum_{l}\mu_{l}\rho_{l}^{v}-\varepsilon_{l},
\end{eqnarray}
where the expression of the vector densities $\rho_{b}^{v}$ and $\rho_{l}^{v}$ are given in~(\ref{densities2}) and (\ref{vector_{density_{leptons}}}), respectively.

\subsection{Density-dependent magnetic field}
\label{B-density-depndence}

We assume that the  magnetic field $B$ in the EOS  depends on the density according to \cite{PRL79:2176:1997,ChJAA3:359:2003,PRC80:065805:2009,JPG36:115204:2009,BJP42:428:2012}
\begin{equation}\label{cmdd}
    B\left(\frac{\rho}{\rho_{0}}\right)=B_{\mathrm{surf}}+B_{0}\left\{ 1-\mathrm{exp}\left[ -\beta\left( \frac{\rho}{\rho_{0}}\right)^{\gamma}  \right] \right\}\,,
\end{equation}
where $\rho=\sum_{b}\rho_{b}^{v}$ is the baryon density, $\rho_{0}$ is the saturation density, $B_{\mathrm{surf}}$ is the magnetic field on the surface of a magnetar, taken equal to $10^{15}$~G in agreement with observational values, and $B_{0}$ is the magnetic field for larger densities. The parameters $\beta$ and $\gamma$ are chosen to reproduce two behaviors of the magnetic field: a fast decay with $\gamma=3.00$ and $\beta=0.02$ and a slow decay with $\gamma=2.00$ and $\beta=0.05$ \cite{PRC89:015805:2014}. According to the discussion in the previous subsection, we use two values for the magnetic field $B_{0}$, namely $10^{17}$~G and $3.1\times10^{18}$~G.

\subsection{On the isotropy of the pressure}

Notice that in the previous subsections we assumed that the matter pressure ($P_m + P_l$) is isotropic in spite of the high values of the magnetic field. As it has been shown in Ref. \cite{JPG41:015203:2014},  the anisotropic effects around $3.1\times10^{18}$~G are small, thus we restrict ourselves to magnetic fields below this value.

However, the purely field-related pressure $P_B \sim  B^2$ may become dominant in the core of the star. In such cases, the total pressure  perpendicular to the magnetic field lines would be significantly larger than the pressure parallel to the field lines. Therefore,  3D or at least 2D stellar structure equations should be used in order to incorporate the effect of the pressure anisotropy. We must stress however, that the magnetic field geometry inside a neutron star can be extremely complex, and depending on its configuration the use of the spherically symmetric stellar structure equations can still be a good enough approximation, as we argue below. 

If we consider, for example, a purely dipole field (purely poloidal field), a 1D stellar structure calculation would be certainly inappropriate for some very high field objects.  But such configuration is unstable and cannot be realised in Nature; in fact, any neutron star with a purely poloidal or a purely toroidal magnetic field is unstable in general relativity (see \cite{Ciolfi2011} and references therein). This strongly supports the idea that any long-lived magnetic field configuration in a NS has to consist of a mixture of poloidal and toroidal field components.
On the other hand, while stellar shapes have long been considered to be oblate due to the effects of centrifugal and/or magnetic forces, this is not necessarily true. For example, it has been found recently that the shape of our Sun is perfectly round, against the common expectation of an oblate shape due to its rotation \cite{Kuhn2012}.

In the specific case of neutron stars, the results found thus far are that purely poloidal magnetic fields make stars oblate (equatorial radius larger than polar radius), while purely toroidal magnetic fields lead stars to become prolate (polar radius larger than equatorial radius). Very recently, it has been shown that equilibrium states of magnetized stars with mixed poloidal$-$toroidal magnetic fields are possible; in particular, it was possible to build neutron stars with twisted-torus configurations in equilibria with any toroidal and poloidal field energy content \cite{Ciolfi2013}. Such poloidal and toroidal magnetic fields act as increasing and decreasing mechanisms for stellar oblateness, respectively \cite{Fujisawa2013}.
An additional finding of Ref. \cite{Ciolfi2013} is that for a fixed polar magnetic field strength, a higher relative content of toroidal field energy ($> 10 \%$) implies in general a much higher total (poloidal and toroidal) magnetic energy inside the star. This means that a  highly magnetized neutron star can harbor internal magnetic fields that are significantly stronger than commonly expected.

In addition, a very complex field is expected from the formation process of magnetized neutron stars, since they are born hot, highly convective and differentially rotating  \cite{Thompson1996,Bonanno2003}. During its early evolution,  neutron star's magnetic fields can be significantly amplified and redistributed by several mechanisms including possibly dynamo action and shear instabilities. In this context, a distribution of magnetic energy with poloidal and toroidal fields close to equipartition appears as a very reasonable candidate for the internal magnetic field configuration. In such a case, since both the toroidal and poloidal components are of  the same order, we may expect that oblateness and prolateness cancel out approximately, leading to stars close to the spherical symmetry. Since a study of stellar pulsations in 2D or 3D is numerically involving, we shall adopt spherical symmetry as a first approximation, and use the much more simple 1D equations. 
Within this approach, we must add to the matter pressure ($P_m + P_l$) given in Sec. \ref{sec:A},  an effective mean magnetic pressure $P_B$ representing the effective isotropic contribution arising from the combined effect of poloidal and toroidal field components of the same order. 
Thus, the total energy density and the total pressure of the system read:
\begin{eqnarray}
\label{et}
\varepsilon & = & \varepsilon_{m}+\varepsilon_{l} + \tfrac{1}{2}  \left[B\left(\frac{\rho}{\rho_{0}} \right)\right]^{2} , \\
P & = & P_{m}+P_{l} + P_B  .\label{et222}
\end{eqnarray}
For finding the effective mean magnetic pressure $P_B$ we shall follow here an averaging procedure similar to the one employed in works that focus on the study of magnetic fluctuations in turbulent fluids \cite{Brandenburg2012,Rogachevskii2007}.  The magnetic stress tensor $\sigma_{ij}$ is given by
\begin{eqnarray}
\sigma_{ij} = - \frac{\langle {\bf B}^2 \rangle}{2}  \delta_{ij} + \langle B_i B_j \rangle , 
\label{stress_tensor}
\end{eqnarray}
where $ \delta_{ij} $ is the Kronecker tensor and the brackets $ \langle \cdots \rangle$  denote the averaging. 
For a completely isotropic distribution of the magnetic field we have $\langle B_i B_j \rangle = \delta_{ij} \, \langle {\bf B}^2 \rangle / 3 $, and the magnetic stress tensor reads
\begin{eqnarray}
\sigma_{ij}= - {\langle {\bf B}^2 \rangle \over 6} \, \delta_{ij} 
\label{I2}
\end{eqnarray}
The effective mean magnetic pressure $P_B$ is related to the magnetic stress tensor by $\sigma_{ij} = - P_B \,
\delta_{ij}$.  Therefore,  we find:
\begin{eqnarray}
P_{B} =   \tfrac{1}{6}  \left[B\left(\frac{\rho}{\rho_{0}} \right)\right]^{2}  .
\label{WA1}
\end{eqnarray}
A similar expression has been used in  \cite{Bednarek2003} for studying the influence of asymmetry on a magnetized proto-neutron star.


\section{Radial Oscillations}

In order to study the radial oscillations of a compact star, we must know first its equilibrium configuration.  Such configuration is perturbed in such a way that the spherical symmetry of the body is not violated. The space-time and fluid perturbations are inserted into the Einstein equations and into the energy, momentum and baryon number conservation equations, assuming a sinusoidal time dependence $\exp{(i \omega t)}$ and retaining only the first-order terms. The result of this procedure is a second order ordinary differential equation for the perturbations \cite{APJ140:417:1964} or alternatively a set of two first order ordinary differential equations, as shown below. In the following we present the explicit form of the equilibrium and oscillation equations employed in the present work. 

\subsection{Equilibrium configuration}
\label{sec3}

We consider that the unperturbed  compact star is totally composed of a perfect fluid. In this case the stress-energy momentum tensor can be expressed as 
\begin{equation}
{T}_{\mu\nu} =(\epsilon + p) u_{\mu}u_{\nu} + p{g}_{\mu\nu},
\end{equation}
where $\epsilon$ and $p$ are the energy density and pressure respectively.

The  background spacetime of the spherically symmetric star,  can be expressed through the line element
\begin{equation}
\label{dsz_tov}
ds^{2}=-e^{ \nu(r)} dt^{2} + e^{ \lambda(r)} dr^{2} + r^{2}(d\theta^{2}+\sin^{2}{\theta}d\phi^{2}),
\end{equation}
where $t,r,\theta,\phi$ are the set of Schwarzschild-like coordinates, and the metric potentials $\nu(r)$ and $\lambda(r)$ are functions of the radial coordinate $r$ only.

The Einstein equations in such a spacetime lead to the following set of stellar structure equations (Tolman-Oppenheimer-Volkoff equations)
\begin{eqnarray}
\frac{dm}{dr} = 4 \pi r^2 \epsilon,  \\
\frac{d\nu}{dr} = - \frac{2}{\epsilon} \frac{dp}{dr} 	\bigg(1 + \frac{p}{\epsilon}\bigg)^{-1},   \\
\frac{dp}{dr} = - \frac{\epsilon m}{r^2}\bigg(1 + \frac{p}{\epsilon}\bigg) \bigg(1 + \frac{4\pi p r^3}{m}\bigg)\bigg(1 - \frac{2m}{r}\bigg)^{-1},
\end{eqnarray}
where $m$ is the gravitational mass inside the radius $r$.

The metric function $\nu$ has the following boundary condition
\begin{equation}
\label{BoundaryConditionMetricFunction}
    \nu(r=R)= \ln \bigg( 1-\frac{2M}{R} \bigg),
\end{equation}
where $R$ is the radius of the star and $M$ its mass. With this condition the metric function $\nu$  will match smoothly to the Schwarzschild metric outside the star. The boundary conditions for $m$ and $p$ are $m(r=0) = 0$ and $p(r=R) = 0$.

\begin{figure}
\includegraphics[angle=0,scale=0.66]{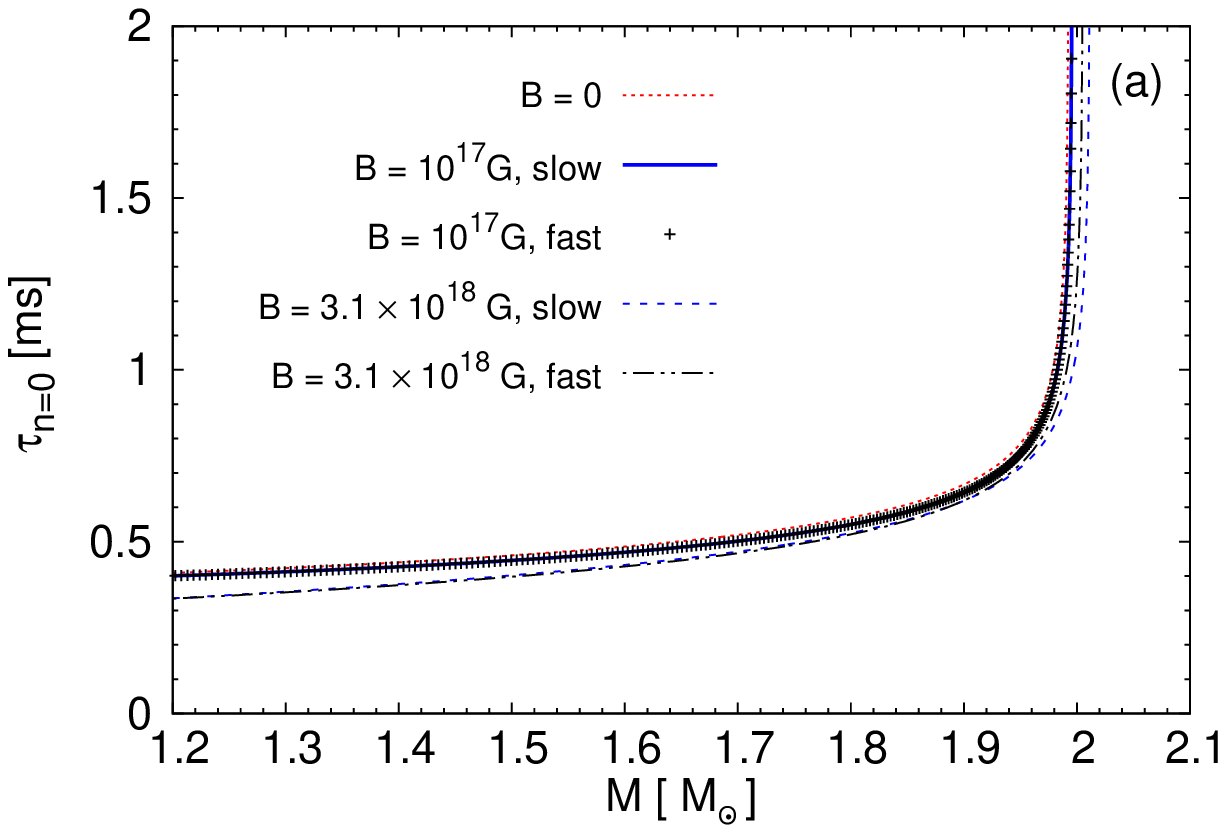}
\includegraphics[angle=0,scale=0.66]{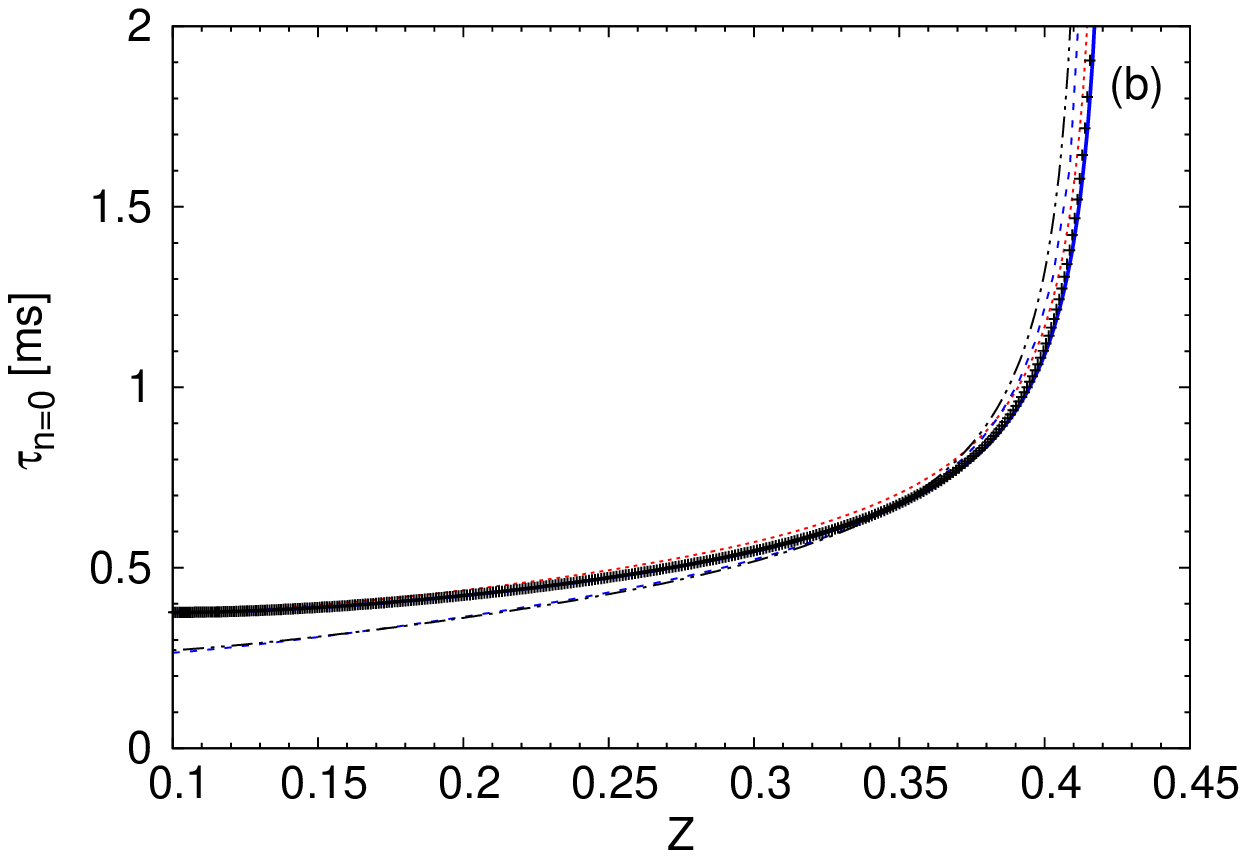} 
\includegraphics[angle=0,scale=0.66]{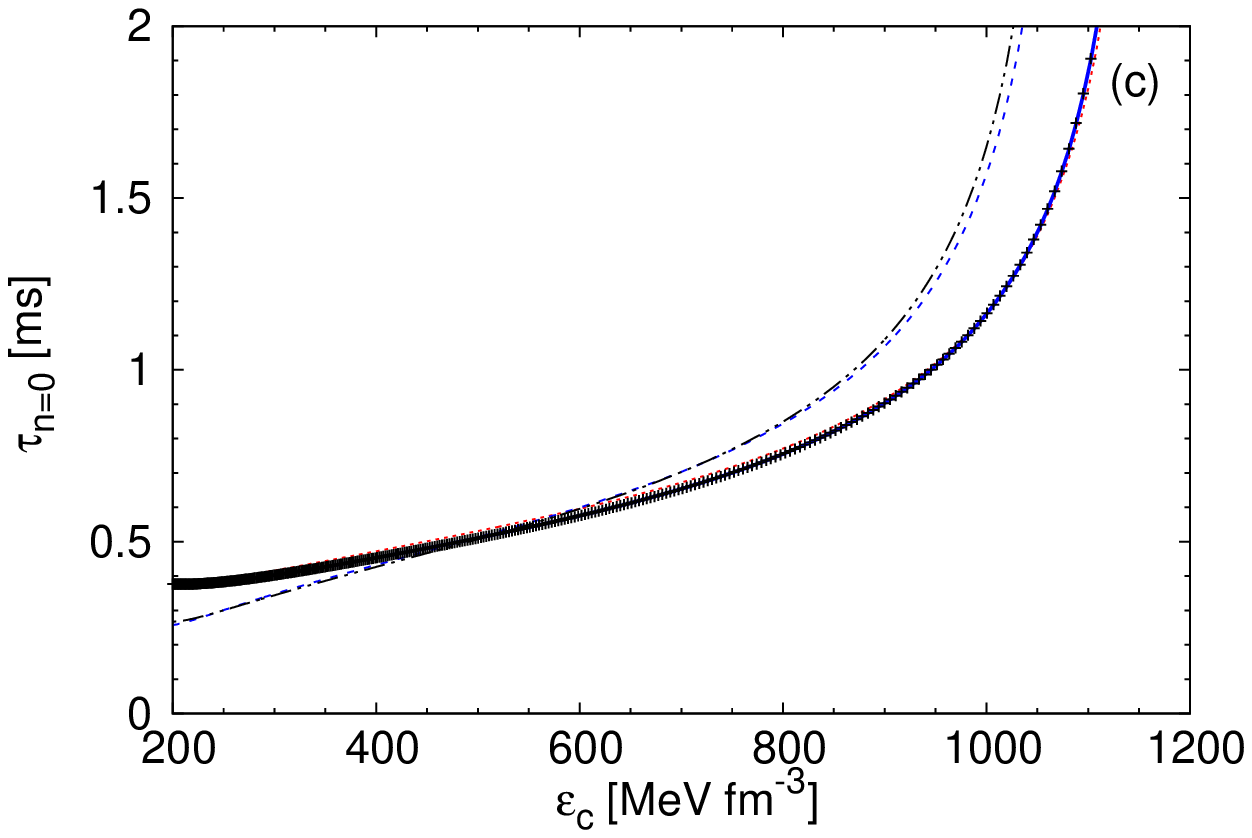}
\caption{The period of the fundamental mode of hadronic stars as a function of (a)  the  mass $M$, (b) the gravitational redshift $Z$, and (c) the central energy density $\epsilon_c$.  In all the figures we set three different values for the magnetic field: B$=0,10^{17}$ G and $3.1 \times 10^{18}$G. We also show the effect of slow and fast decay in the magnetic field profile. }
\label{fig1}
\end{figure}

\begin{figure}
\includegraphics[angle=0,scale=0.66]{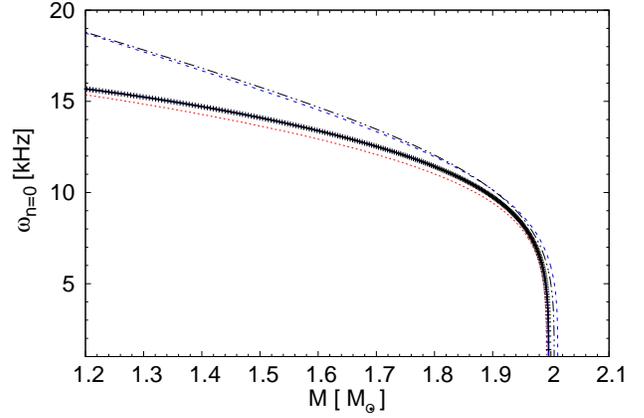}
\caption{The frequency of the fundamental mode ($\omega_{n=0}=2\pi/ \tau_{n=0}$) as a function of the neutron star mass. The values for the magnetic field as well as the slow and fast decays in the magnetic field profile are the same as given in Fig. \ref{fig1}. }
\label{fig2}
\end{figure}

\subsection{Oscillation equations}
\label{sec4}

Several forms of the oscillation equation have been presented in the literature (for more details the reader is referred to \cite{PRD82:063006:2010}). In this work, we use the set of equations of Gondek et al. \cite{AA325:217:1997} and adopt $G = c = 1$. The equations read 
\begin{eqnarray} \label{ecuacionparaXI}
\frac{d\xi}{dr} &=& -\frac{1}{r}\bigg(3\xi+\frac{\Delta p}{\Gamma p}\bigg)-\frac{dp}{dr}\frac{\xi}{(p+\epsilon)},   \\
 \label{ecuacionparaP}
\frac{d\Delta p}{dr} &=& \xi \bigg\{\omega^{2}e^{\lambda-\nu}(p+\epsilon)r-4\frac{dp}{dr} \bigg \}  \nonumber \\  
&& +\xi\bigg \{\bigg(\frac{dp}{dr}\bigg)^{2}\frac{r}{(p+\epsilon)}-8\pi e^{\lambda}(p+\epsilon)pr \bigg \}   \nonumber \\ 
&& +\Delta p\bigg \{\frac{dp}{dr}\frac{1}{(p+\epsilon)} - 4\pi(p+\epsilon)r e^{\lambda}\bigg\},
\end{eqnarray}
where $\omega$ is the eigenfrequency and the quantities $\xi \equiv \Delta r / r$ and $\Delta p$ are assumed to have a harmonic time dependence $\varpropto e^{i\omega t}$.

To solve equations (\ref{ecuacionparaXI}) and (\ref{ecuacionparaP}) one needs two boundary conditions.
The  condition of regularity at the centre ($r=0$) can be written as \cite{AA260:250:1992,AA325:217:1997,AA344:117:1999}
\begin{equation}\label{DeltaP}
(\Delta p)_{center}=-3(\xi \Gamma p)_{center}.
\end{equation}
where the eigenfunctions are normalized in order to have $\xi(0)=1$.  The second boundary condition, expresses the fact that the Lagrangian perturbation in the pressure at the  stellar surface is zero, thus:
\begin{equation}
\label{PenSuperficie}
(\Delta p)_{surface}=0.
\end{equation}

To solve numerically the oscillation equations  we employ a shooting method in order to fulfil  the required boundary conditions. For more details on the method see Ref. \cite{PRD82:063006:2010}.

\section{Results and conclusions}

In this section we analyse the effect that a strong magnetic field could produce on  the fundamental mode of the radial oscillations of hadronic stars. As mentioned before, we consider that the magnetic field decays with the density following the fast and slow profiles presented  in Section \ref{B-density-depndence}. All the models for hadronic stars investigated in the present work have a maximum mass in agreement with the recent observation of the pulsars PSR J1614-2230 with $M = (1.97 \pm 0.04) M_{\odot}$ \cite{Demorest2010} and  PSR J0348-0432 with $M = (2.01 \pm 0.04) M_{\odot}$ \cite{Antoniadis2013}.

In Fig. \ref{fig1} we see that a magnetic field profile with $B_0 = 10^{17}$ G, produces very small changes  on the oscillation period of the fundamental mode with respect to an unmagnetized star, for both slow and fast decays. This can be explained by the small effect that such magnetic field intensity has on the equation of state. 
In contrast, when $B_0 = 3.1 \times 10^{18}$ G is selected, there is a clear change in the oscillation period.  As a function of the stellar mass, the curves fall below and to the right of the curves for weaker fields. Notice that for large mass objects the period changes because of the  shift of the curves due to the increase of the maximum stellar mass. For smaller masses, the curves for  $B_0 = 3.1 \times 10^{18}$ G are also significantly different with respect to the unmagnetized case; e.g. for a neutron star with $1.4 M_{\odot}$  the period is around $20 \%$ smaller, and the difference increases for less massive stars. For completeness we present also  the oscillation period  as a function of the gravitational redshift  (see middle panel of Fig. \ref{fig1}) and  as a function of the central mass-energy density  (lower panel of Fig. \ref{fig1}). 

The effect of strong magnetic fields is more apparent in the frequency of the fundamental mode as can be seen in Fig. \ref{fig2}. The oscillation frequency for $B_0 = 10^{17}$ G is slightly above the one of an unmagnetized object of the same mass, and there is almost no difference between the fast and the slow decaying profiles of the magnetic field. However, if $B_0 = 3.1 \times 10^{18}$ G the oscillation frequencies are clearly larger than for an unmagnetized star of the same mass. For example, for a star with $1.4 M_{\odot}$  the frequency is around $20 \%$ larger and for a  $1.7 M_{\odot}$ star it is around $10 \%$ larger. The difference between the fast and the slow decaying profiles of $B$ is very small.

As stated before, purely radial modes do not emit gravitational waves and consequently they are essentially damped by the bulk viscosity, 
originated from the re-establishment of chemical equilibrium when a fluid element of the star is compressed and rarified during pulsations. 
Unfortunately, there is great uncertainty about the amount of viscosity inside neutron stars since it depends sensitively on the composition of matter which is uncertain beyond few times the nuclear saturation density 
\cite{Jones2001,Lindblom2002,Haensel2002,Chatterjee2007,Jha2010}.  If the damping time due to viscous forces is long enough, radial pulsations in magnetars can leave an imprint in the microstucture of magnetar flare lightcurves opening a new window for the study of highly magnetized ultradense matter.

\acknowledgments

L. B. Castro thanks CNPq, Brazil, Grants No 455719/2014-4 (Universal) and No 304105/2014-7 (PQ) for partial support. C. V\'asquez Flores acknowledges the financial support of CAPES, Brazil. G. Lugones acknowledges  FAPESP  and CNPq for financial support. We acknowledge an unknown referee for valuable comments.

\end{document}